\documentclass{article}
\usepackage{spconf,amsmath,graphicx}

\usepackage{xcolor}
\usepackage{multirow}
\usepackage{makecell}
\usepackage{booktabs}
\usepackage{amsfonts}
\usepackage{subcaption}
\usepackage{enumitem}
\usepackage{url}

\newcommand{\tb}[1]{\textbf{#1}}

\title{MultiSV: dataset for far-field multi-channel speaker verification}

\name{Ladislav Mo\v{s}ner, Old\v{r}ich Plchot, Luk\'{a}\v{s} Burget, Jan ``Honza'' \v{C}ernock\'{y}\thanks{The work was supported by Czech Ministry of Interior project No. VJ01010108 "ROZKAZ", Czech National Science Foundation (GACR) project NEUREM3 No. 19-26934X, Czech Ministry of Education, Youth and Sports from project no. LTAIN19087 "Multi-linguality in speech technologies", and Horizon 2020 Marie Sklodowska-Curie grant ESPERANTO, No. 101007666. Computing on IT4I supercomputer was supported by the Czech Ministry of Education, Youth and Sports from the Large Infrastructures for Research, Experimental Development and Innovations project "e-Infrastructure CZ – LM2018140".}}
\address{Brno University of Technology, Faculty of Information Technology, Speech@FIT, Czechia}

\begin{document}
\ninept
\maketitle
\begin{abstract}
Motivated by unconsolidated data situation and the lack of a standard benchmark in the field, we complement our previous efforts and present a comprehensive corpus designed for training and evaluating text-independent multi-channel speaker verification systems. It can be readily used also for experiments with dereverberation, denoising, and speech enhancement. We tackled the ever-present problem of the lack of multi-channel training data by utilizing data simulation on top of clean parts of the Voxceleb dataset. The development and evaluation trials are based on a retransmitted Voices Obscured in Complex Environmental Settings (VOiCES) corpus, which we modified to provide multi-channel trials.  We publish full recipes that create the dataset from public sources as the MultiSV corpus, and we provide results with two of our multi-channel speaker verification systems with neural network-based beamforming based either on predicting ideal binary masks or the more recent Conv-TasNet.
\end{abstract}
\begin{keywords}
Multi-channel, speaker verification, MultiSV, dataset, beamforming
\end{keywords}
%

%%%%%%%%%%%%%%%%%%%%%%%%%
\section{Introduction}
\label{sec:introduction}
%%%%%%%%%%%%%%%%%%%%%%%%%
Single-channel text-independent speaker verification (SV) has traditionally attracted more research attention than multi-channel SV. Therefore, we aim at fostering multi-channel SV by designing training and evaluation MultiSV corpus. The need for progress in the single-channel field is rational and natural. It has been promoted by numerous NIST SRE evaluations (among others) to advance speaker verification in telephony \cite{nist_evals}. Datasets provided for the aforementioned evaluations are, however, available only to participants. Therefore, the release of the Voxceleb 1 \cite{Voxceleb1} and 2  \cite{Voxceleb2} datasets was a great deed for the research community. They provide a diverse and extensive collection of utterances of more than 7,000 speakers. Voxceleb also provides an evaluation protocol comprising multiple versions: O (original), H (hard), and E (extended). The need for such a dataset has been affirmed by multiple recent challenges, where it served as a training dataset \cite{nagrani2020voxsrc}. It is usually extended by augmentations for increased performance and robustness of current state-of-the-art embedding extractors. Such data-hungry models are inspired by the seminal work on x-vectors \cite{Snyder2017}.

Popular home devices, televisions, and smart speakers usually contain microphone arrays. Due to different locations on the device, they can provide spatial information that is useful for various speech processing tasks \cite{chime5_challenge}. SV is one of the tasks that such devices need to support to provide personalized responses, actions, and content. While single-channel data collection is much easier, gathering a sufficient multi-channel dataset (for research purposes in a non-commercial domain) is prohibitively costly and time-demanding. This is likely the reason for the non-existence of a large multi-channel corpus that could be utilized for speaker verification purposes.

Previous studies on multi-channel SV often adopted some single-channel corpus (such as dataset by SiTEC used in \cite{Yang2019_mvdr_and_xvec} or AISHELL-ASR0009-[ZH-CN] in \cite{Cai2019}) with room-corrupted signals created by simulation. In some works \cite{Taherian2019, Taherian2020}, publicly available, albeit non-free datasets, such as single-channel NIST SRE 2008 data, are utilized for training. The mentioned inconsistency points out the lack of common multi-microphone data for SV experiments and difficulties with reproducibility and comparison.

Over the years, a few data collecting or assembling activities have emerged. One of them is the HI-MIA dataset \cite{hi_mia}. It is unique in that it comprises recordings of 340 real speakers (not retransmitted speech). Each speaker uttered 160 (English and Chinese) wake words and was recorded by six 16-channel microphone arrays and one close-talking microphone. The corpus contains recordings both with and without background noise. It is composed of two sub-datasets:  AISHELL-wakeup and AISHELL-2019B-eval. The former was used in \cite{hi_mia} for fine-tuning and the latter for evaluation. AISHELL-wakeup was not used for embedding extractor training from scratch as it contains only 16 hours per microphone (which is not enough for current models). As suggested by the name, the dataset was designed for text-dependent SV.

We note that Task 2 of the Far-Field Speaker Verification Challenge (FFSVC) \cite{ffsvc} dealt with far-field text-independent SV. It consists of Mandarin data and provides a training set comprising microphone array recordings of 120 speakers. For evaluation, the enrollment segments were recorded with a close-proximity cell phone. Test segments were captured with microphone arrays (4 channels).

Libri-adhoc40 \cite{libri_adhoc40} is a recent collection of retransmitted precisely synchronized audio. It provides speech of 331 speakers from the LibriSpeech corpus \cite{librispeech}. Uniquely, 40 equal microphones were used for collection where each recorded 110 hours of speech. The placement of microphones and speakers for the training part (train-clean-100) differed from the settings for the development/evaluation part (dev-clean, test-clean). The collection took place in a large reverberant room (with RT60 900~ms). For reference, the same data was also recorded in an anechoic chamber. Compared to HI-MIA, no additive noise was replayed. Although \cite{libri_adhoc40} presents ASR results, the dataset can be used for text-independent SV as shown in \cite{sv_adhoc40}.

Some works tried to re-use existing data to prepare SV trials. CHiME-5 challenge \cite{chime5_challenge} surprised by a great difficulty of the real data collected during numerous parties of friends. The appeal of this data led to the CHiME-5 speaker recognition benchmark \cite{Garcia-Romero2019}. It reused recordings from 18 (out of 20) parties of 39 speakers. It aims at providing means to compare various scenarios: single speaker vs. multiple speakers, close-talking vs. far-field microphones, and single microphone vs microphone arrays. Unfortunately, despite its attractivity, it has never been released to the public.

In \cite{Odyssey2020}, we presented trial definitions for text-independent  multi-channel speaker evaluation based on a retransmitted VOiCES corpus \cite{voices_corpus}. It alters the single-channel VOiCES challenge \cite{voices_challenge} trials by grouping test microphones to ad-hoc arrays.

We present a comprehensive MultiSV corpus that complements previous works and aims at fostering research in multi-channel text-independent SV, enabling and easing comparison. The following points summarize our motivation and contributions:
\begin{itemize}[leftmargin=*]
    \setlength\itemsep{-0.5mm}
    \item Even though HI-MIA contains recordings of real speech (not retransmitted), it has been designed for text-dependent SV. Therefore, it is not suitable for text-independent SV. Even though it provides training data recorded over many microphones, it contains only 16 hours per microphone. In contrast with HI-MIA, MultiSV targets text-independent SV and provides a training set of simulated 4-microphone arrays with approximately 77 hours per microphone.
    \item Libri-adhoc40 is a large corpus recorded with 40 microphones. The data are affected by reverberation, but it does not contain background noise which is likely to occur in far-field scenarios. MutliSV data contain both background noise and reverberation.
    \item CHiME-5 speaker recognition benchmark comprises only evaluation data (training data is not a part of it). More importantly, it has never been released to the public. MultiSV corpus is intentionally based on publicly available data. We released lists and code for preparation from scratch at \url{https://github.com/Lamomal/MultiSV}.
    \item Along with speaker labels for training data, we provide clean (or reverberant) speech and clean (or reverberant) noise references. Therefore, apart from SV, MultiSV can be used also for dereverberation/denoising/speech enhancement model training.
    \item We extend \cite{Odyssey2020}, which introduced a single-channel enrollment and multi-channel test protocol, by providing multiple conditions with single-channel or multi-channel enrollment segments (keeping multi-channel test recordings).
\end{itemize}

%%%%%%%%%%%%%%%%%%%%%%%%%
\section{MultiSV corpus}
\label{sec:multisv}
%%%%%%%%%%%%%%%%%%%%%%%%%

\subsection{Training data}
%%%%

For the expensiveness of multi-channel data collection, we opted for a compromise -- simulation. Simulated training data are common in multichannel SV literature \cite{Cai2019, Yang2019_mvdr_and_xvec} but also in other fields of speech processing where clean references are needed. We decided to base the data on Voxceleb 2 dev since it is often used for training of embedding extractors. Moreover, we do not constrain a variety of possible evaluation data by taking speakers from different datasets (such as LibriSpeech in Libri-adhoc40 \cite{libri_adhoc40}).

Recordings from the development part of Voxceleb 2 contain various channels and noise levels. Since they are supposed to be used as a source of speech in data simulation, clean signals are preferred. To this end, we performed a pre-selection utilizing SNR estimation. Only recordings with SNRs exceeding 20\,dB were preserved. The final selection was split into the training and cross-validation parts. The training part contains speech of 1,000 speakers with an even gender split (500 female and 500 male voices). For each speaker, the set contains at most 50 utterances. All recordings amount to 72.26 hours of audio. The cross-validation set comprises 90 speakers (45 females, 45 males), each represented by  35 utterances at most. The total duration of this part is 4,68 hours. The average duration of utterances in both parts is 6.2\,s. Statistics of the training set are shown in Table \ref{tab:train_set_stats}.

To mimic indoor conditions, we performed a simulation considering a point source of speech and a point source of noise. Noises likely to occur indoors were selected from three sources:
\begin{itemize}[leftmargin=*]
\itemsep0em
\item \tb{Music} (66.3\,h) -- FMA \emph{small} dataset \cite{fma_dataset} where we excluded recordings that are present in the music part of MUSAN \cite{musan2015} as these are used in the evaluation data,
\item \tb{MUSAN noises} (5.0\,h) -- 80\% of the ``noise" part of the MUSAN dataset (i.e. without music and babble),
\item \tb{Freesound.org and self-recorded noises} (20.1\,h) -- real fan, HVAC, shop, crowd, library, office, and dishes sounds.
\end{itemize}
In order to obtain reverberant speech and noise, we performed room impulse response (RIR) generation by image source method (ISM) \cite{Allen_ISM}. We simulated box-shaped rooms with 4-microphone uniform linear arrays (ULA). Length of ULAs ranged from 10 cm to 2 m. Reverberation time RT60 was sampled uniformly from a range of [0.3, 0.9] s. Resulting RIRs were convolved with speech and noise sources. Mixing SNRs were uniformly drawn from [3, 20] dB.

For multi-channel SV models where a single-channel pre-training is possible (and/or required), we propose to pair MultiSV with Voxceleb 2 dev. It will enable a fair comparison.

\begin{table}[tb]
    \centering
    \caption{Per-microphone statistics of the MultiSV training set.}
    \begin{tabular}{l c c}
        \toprule
            & \tb{Training} & \tb{Cross-validation} \\
        \midrule
            \tb{Duration [h]} & 72.62 & 4.68 \\
            \tb{Speakers} & 500 + 500 & 45 + 45 \\
            \tb{Files} & 41,854 & 2,720 \\
        \bottomrule
    \end{tabular}
    \label{tab:train_set_stats}
    \vspace{-0.5cm}
\end{table}

\subsection{Development and evaluation data}
\label{sec:dev_eval_data}
%%%%

We base development and evaluation trials on those defined for the VOiCES challenge \cite{voices_challenge} keeping the same utterances. Data for the challenge \cite{voices_corpus} was obtained by retransmitting speech from LibriSpeech together with babble, television, music, or none (diffuse background) noises. The audio was simultaneously recorded by multiple studio-quality and lapel microphones. Despite the availability of multi-microphone recordings, the VOiCES challenge used single-channel enrollment and test segments. On the contrary, our trials contain multi-channel test recordings. They were obtained either by grouping four far-field VOiCES recordings with the same content (forming ad-hoc arrays) or by simulation of microphone arrays. To support various use cases, we devised multiple conditions based on properties of the enrollment segments:
\begin{itemize}[leftmargin=*]
    \item \tb{CE} (clean enrollment): comprises the same trials as \cite{Odyssey2020} but enrollment recordings are always clean. They contain the same utterances but the utterances were selected from the LibriSpeech source audio (that was used for retransmission),
    \item \tb{SRE} (single-channel retransmitted enrollment): corresponds exactly to \cite{Odyssey2020}: single-channel enrollment segments are either retransmitted (dev) or a combination of retransmitted and clean recordings (eval),
    \item \tb{MRE} (multi-channel retransmitted enrollment): similar to SRE but enrollment segments are 4-channel recordings from quiet rooms (that corresponds to 4-microphone ad-hoc arrays in a reverberant room),
    \item \tb{MRE hard} similar to MRE but with the noise source playing in the background.
\end{itemize}

MultiSV defines development and evaluation sets containing different lists of verification trials. The retransmitted development set (\emph{dev retr}) is derived from the original VOiCES development set as follows: In the original VOiCES trial list, we identified groups of 4 trials sharing exactly the same enrollment recording (retransmitted and recorded by the same microphone) and the same test utterance recorded by 4 different microphones. Each group of such 4 trials was reduced to a single trial with multi-channel test audio. The retransmitted evaluation set (\emph{eval retr v1}) was constructed analogically from the original VOiCES evaluation set.
The original signals were recorded by microphones placed across a large L-shaped room (room 4 \cite{Odyssey2020}). Sensors placed right next to a distractor speaker (mic. 12), fridge (mic. 19), or behind a corner far from the source speaker (mic. 6) are included as well\footnote{A floor plan of the room 4 can be found at \url{https://iqtlabs.github.io/voices/rooms/}. We found microphones 6, 12, 17, and 19 problematic. They were replaced by microphones 1, 2, 3, and 7.}. Signals of such microphones may have negative SNRs and be confusing for single-channel enhancement models: Poor predictions for such microphones in an array can lead to erroneous results. Due to the described properties reflected in \emph{eval retr v1}, we defined a second version (\emph{eval retr v2}). The only difference compared to v1 is that the problematic microphones were replaced by those that are close to non-problematic ones.

The simulated sets of multi-channel test segments use the same source speech as retransmitted sets. Therefore, the resulting development (\emph{dev simu}) and evaluation (\emph{eval simu}) definitions are equal to the retransmitted ones in terms of speakers, utterances, and the number of trials. The difference is that the 4-microphone arrays were simulated. The domains of the training distractors and those used here are the same (but the recordings are different):
\begin{itemize}[leftmargin=*]
  \itemsep0em
\item \tb{Music} (4.2\,h $+$ 4.2\,h) -- a subset of MUSAN music equally split between the development and evaluation sets, 
\item \tb{MUSAN noises} (0.6\,h $+$ 0.6\,h) -- 10\% and 10\% of the ``noise" part of MUSAN, 
\item \tb{Freesound.org and self-recorded noises} (1.8\,h $+$ 2.2\,h) -- similar to the training noises.
\end{itemize}
RT60 reverberation times of ISM-generated RIRs were uniformly drawn from the interval [0.3, 0.9] s. Mixing SNRs were uniformly drawn from [3, 20] dB similarly as for the training set.

The development set comprises 196 speakers and 996,448 trials (with 5,024 target ones). The evaluation set comprises 100 different speakers and 973,929 trials (with 9,939 target ones).

For multi-channel SV experiments, we set the prior probability of a target trial $P_{tar}$ to 0.01 when reporting MinDCF. It was used in the VOiCES challenge \cite{voices_challenge} and for current systems, given this dataset, it constitutes an operating point providing statistically reliable results. It can be observed from Figure~\ref{fig:dcf} that this operating point sits reliably on the right side of the FA DR30 point which marks the place where systems make just 30 false alarms and setting an operating point close or to the left of FA DR30 would produce unreliable results (for more details see appendix B of \cite{brummer2010measuring}).

%%%%%%%%%%%%%%%%%%%%%%%%%
\section{Baseline models}
\label{sec:baseline}
%%%%%%%%%%%%%%%%%%%%%%%%%
We provide two baseline systems. Both are multi-channel embedding extraction models composed of a multi-channel speech enhancement and a single-channel x-vector-like speaker embedding extraction. The first system is an updated version of \cite{ASRU2019}. The second one was proposed in \cite{conv_tasnet_SV}. They utilize the same embedding extractor and minimum variance distortionless response (MVDR) beamformer \cite{Capon_MVDR}. What differentiates them is the front-end processing. Both models are trained on the MultiSV training part.

\subsection{Beamforming-based enhancement models}
\label{sec:bf_enhancement}
%%%%
Multi-channel enhancement of the first model comprises \emph{mask-based} beamforming \cite{Erdogan2016, nn-gev}. It utilizes a neural network that estimates per-channel speech and noise masks given the magnitude spectra. Masks are then combined by averaging and used for spatial covariance matrix (SCM) estimation for beamforming. The mask estimator is trained to optimize a binary cross-entropy between outputs and ideal binary masks \cite{wang2018transact}. The model comprises a long short-term memory (LSTM) layer (providing outputs of the same dimensionality as inputs, i.e. 513 which corresponds to the number of frequency bins) followed by two fully connected (FC) layers with 513 neurons and two parallel FC layers predicting masks. Since the first baseline is based on a mask-predicting front-end, we refer to it as \emph{mask predictor} in Tables \ref{tab:results_retrans} and \ref{tab:results_simu}.

The second system comprises a time-domain enhancement model. It splits an input corrupted speech into corresponding speech and noise time-domain components. These outputs (for all channels) are used to provide ratio masks after transformation by STFT. Product-pooled masks serve the same purpose as in the previous approach. The model is trained with an SNR loss. The architecture is a down-scaled Conv-TasNet \cite{conv_tasnet} to reduce the number of trainable parameters. Details on it are available in \cite{conv_tasnet_SV}. Following the name of the speech enhancer, we refer to the second baseline as \emph{Conv-TasNet}-based model.

\subsection{Single-channel embedding extractor}
\label{sec:embd_extract}
%%%%
Speaker embedding extractor is a network predicting utterance-level representations (vectors) of speakers given the beamformer output that is transformed to 40-dimensional log-Mel filter bank energy features (fbanks). Frame length and shift are 25 and 10\,ms, respectively.

Architecture of the extractor is based on ResNet34 \cite{ResNet}. Since it is tailored towards audio processing, the input has only one channel processed by a convolutional layer with a 3 $\times$ 3 kernel. Stages of the network have 64, 128, 256, and 256 channels in that order. The last residual block is followed by mean and standard deviation pooling. The statistics are projected to a 256-dimensional embedding. An additive margin (AM) softmax \cite{AM_softmax} is optimized during training while increasing the scale hyperparameter up to 0.2.

We note that the results obtained with the \emph{mask predictor} baseline are not directly comparable with results in \cite{ASRU2019} for the following reasons: The embedding extractor was switched from a simpler time delay NN architecture (TDNN) with cross-entropy to ResNet with additive margin loss (AM); we employ simpler cosine-similarity scoring instead of PLDA; the training data has changed as well.

For robustness of the embedding extractor, we train it on the full Voxceleb 2 dev. Compared to the clean subset (utilized for MutliSV simulation), it contains approximately 5 $\times$ more speakers. On top of that, we make use of Kaldi-prepared augmentations.

\begin{table*}[th]
    \centering
    \caption{Speaker verification results of baseline models obtained on all the conditions with retransmitted data provided by MultiSV.}
    \resizebox{\linewidth}{!}{%
    \begin{tabular}{l c c c c c c | c c c c c c}
        \toprule
         \tb{Front-end} & \multicolumn{6}{c}{\tb{Mask predictor}} & \multicolumn{6}{c}{\tb{Conv-TasNet-based}} \\
        \midrule
         \multirow{2}{*}{\tb{Condition}} &  \multicolumn{2}{c}{\tb{dev retr}} & \multicolumn{2}{c}{\tb{eval retr v1}} &
         \multicolumn{2}{c}{\tb{eval retr v2}} &
         \multicolumn{2}{c}{\tb{dev retr}} & \multicolumn{2}{c}{\tb{eval retr v1}} &
         \multicolumn{2}{c}{\tb{eval retr v2}} \\
         & \tb{EER [\%]} & \tb{MinDCF} & \tb{EER [\%]} & \tb{MinDCF} & \tb{EER [\%]} & \tb{MinDCF} & \tb{EER [\%]} & \tb{MinDCF} & \tb{EER [\%]} & \tb{MinDCF} & \tb{EER [\%]} & \tb{MinDCF}\\
        \midrule
            \tb{CE} & 0.92 & 0.111 & 4.25 & 0.326 & 1.81 & 0.186 & 0.90 & 0.104 & 4.24 & 0.330 & 1.86 & 0.178 \\
            \tb{SRE} & 0.98 & 0.124 & 4.47 & 0.354 & 2.04 & 0.197 & 0.97 & 0.119 & 4.39 & 0.346 & 2.08 & 0.188 \\
            \tb{MRE} & 1.22 & 0.127 & 3.91 & 0.355 & 2.43 & 0.265 & 1.26 & 0.119 & 3.71 & 0.364 & 2.25 & 0.260\\
            \tb{MRE hard} & 1.35 & 0.137 & 5.37 & 0.518 & 4.90 & 0.440 & 1.31 & 0.129 & 4.61 & 0.482 & 3.99 & 0.387 \\
        \bottomrule
    \end{tabular}
    }
    \label{tab:results_retrans}
\end{table*}

\begin{table}[th]
    \centering
    \caption{Speaker verification results of baseline models obtained on trials employing simulated multi-channel test segments.}
    \resizebox{\linewidth}{!}{%
    \begin{tabular}{l c c c c}
        \toprule
            \multirow{2}{*}{\shortstack{\tb{Front-end}\\ \tb{base}}} &
            \multicolumn{2}{c}{\tb{dev simu}} &
            \multicolumn{2}{c}{\tb{eval simu}} \\
            & \tb{EER [\%]} & \tb{MinDCF} & \tb{EER [\%]} & \tb{MinDCF} \\
        \midrule
            \tb{Mask predictor} & 1.41 & 0.162 & 2.02 & 0.195 \\
            \tb{Conv-TasNet} & 1.17 & 0.147 & 1.91 & 0.176 \\
        \bottomrule
    \end{tabular}
    }
    \label{tab:results_simu}
    \vspace{-0.5cm}
\end{table}

%%%%%%%%%%%%%%%%%%%%%%%%%
\section{Experiments}
\label{sec:experiments}
%%%%%%%%%%%%%%%%%%%%%%%%%
We provide baseline results of the described models for all the provided trial lists. We present both equal error rate (EER [\%]) and minimum detection cost (MinDCF).  For our evaluation to be reliable we made sure there is no overlap between speakers and noise recordings in the training and test parts. Utterances from both parts also differ in terms of speaking style (spontaneous vs. read speech).

\begin{figure*}[tbh]
  \centering
  \begin{subfigure}[t]{0.33\linewidth}
    \centering
    \includegraphics[width=\linewidth,trim={1.5cm 8cm 1.5cm 8cm}]{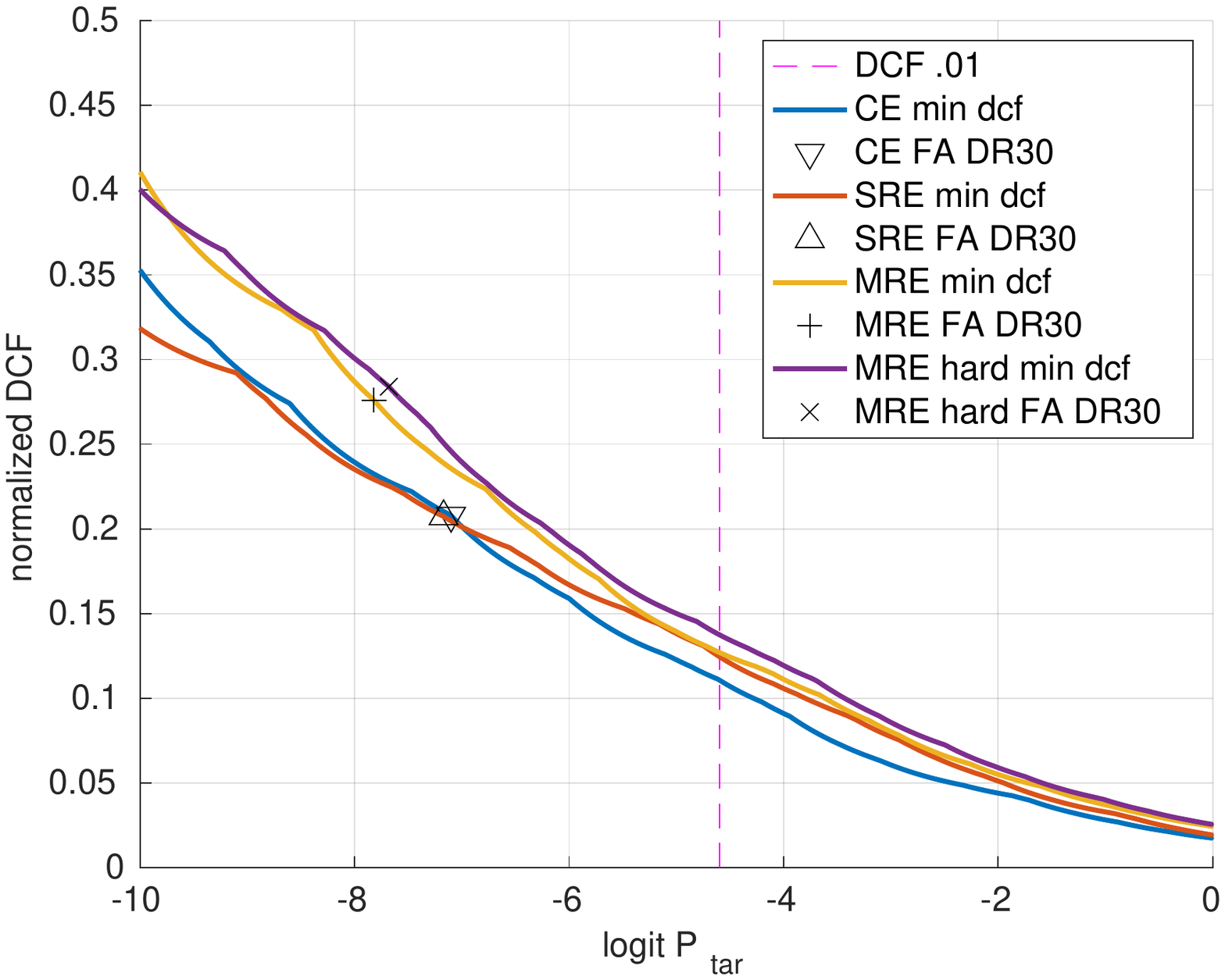}
    \caption{dev retr}
  \end{subfigure}
    % \centering
  \begin{subfigure}[t]{0.33\linewidth}
    \centering
    \includegraphics[width=\linewidth,trim={1.5cm 8cm 1.5cm 8cm}]{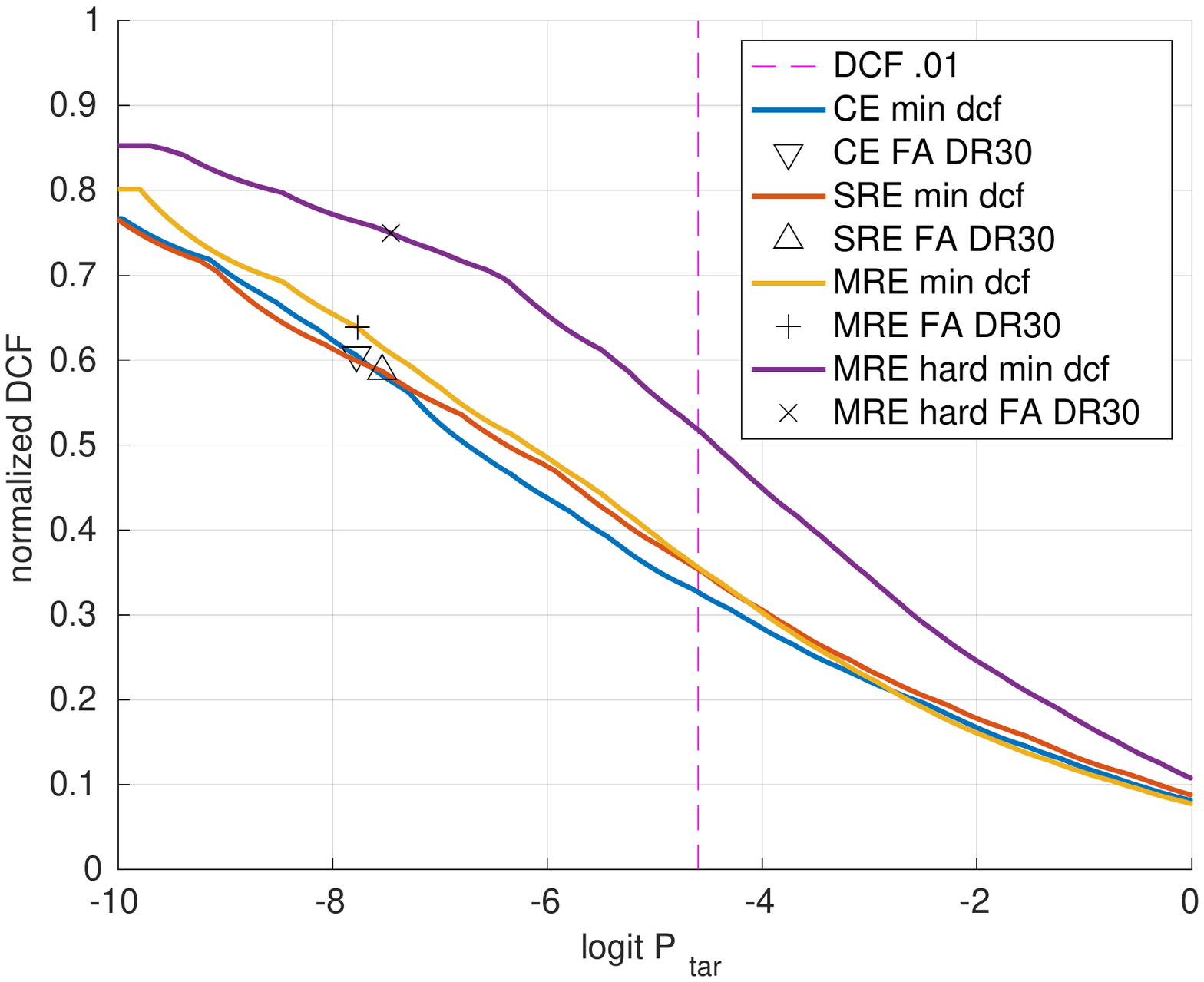}
    \caption{eval retr v1}
  \end{subfigure}
  \begin{subfigure}[t]{0.33\linewidth}
    \centering
    \includegraphics[width=\linewidth,trim={1.5cm 8cm 1.5cm 8cm}]{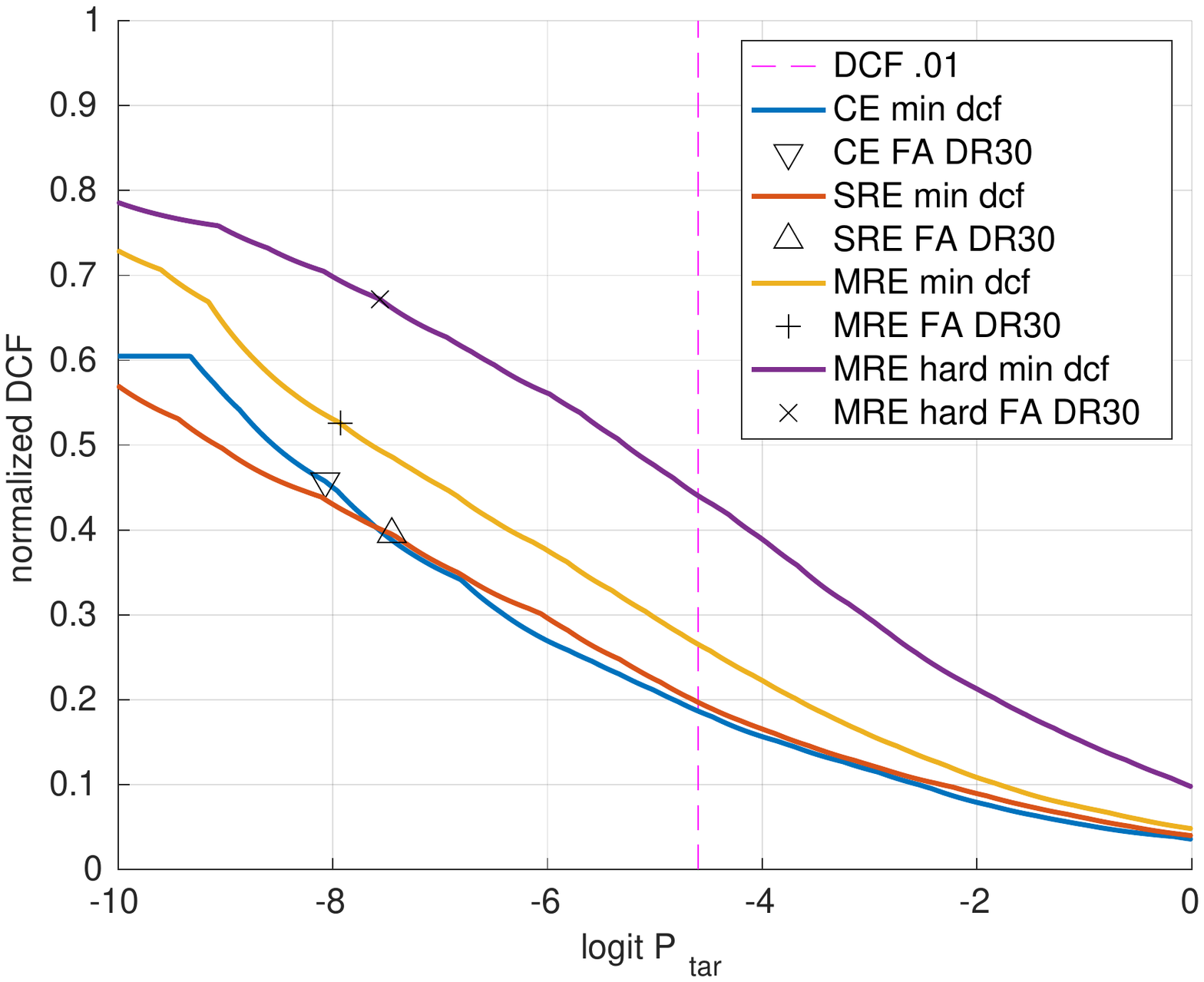}
    \caption{eval retr v2}
  \end{subfigure}
  \caption{Normalized MinDCF as a function of effective prior (a logit of $P_{tar}$) for the mask-predicting baseline. FA DR30 refers to the point to the left of which there are fewer than 30 false-alarms. The vertical magenta line represents the operating point where $P_{tar}=0.01$.}
  \label{fig:dcf}
\end{figure*}

EER and MinDCF obtained for simulated data, \emph{dev simu} and \emph{eval simu}, are shown in Table \ref{tab:results_simu}. Despite the similarity of data corruption in the development and evaluation parts, the results for \emph{eval simu} are worse. This is caused by more difficult enrollment recordings as confirmed by \cite{BUT_VOICES_submission}.

A greater difficulty of the evaluation trial sets (as compared to development trials) is also observable from results in Table \ref{tab:results_retrans} with retransmitted test segments.  
Generally, the \emph{v1} evaluation tends to be more difficult than \emph{v2} due to increased corruption of speech signals caused by the location of used microphones. \emph{V1} is more suitable for meeting-like scenarios, while \emph{v2} is tailored towards investigation-like scenarios with hidden microphones covering large space.

It might be expected that the MRE condition is easier than the SRE. This is not true in our results. In fact, following \cite{voices_challenge}, SRE employs only a limited number of high-quality microphones for enrollment. MRE, on the other hand, comprises recordings from more microphones, including low-quality ones.

In Figure \ref{fig:dcf}, MinDCF is presented as a function of an effective prior for the system based on masks prediction. We observe the same trends for \emph{dev retr} and \emph{eval retr v2} over a wide range of reasonable operating points: the CE condition is the easiest one while MRE hard poses the greatest challenge. Gaps between the curves reflect acoustic profiles of the recording rooms: Evaluation data was retransmitted in a larger and more reverberant room.

We observe clear degradation when moving from \emph{eval retr v2} to \emph{v1}. Due to the greatest mismatch between enrollment and test segments among conditions, the CE and SRE conditions are affected the most. As a result, the corresponding curves get close to (or overlap with) the MRE curve. Even though we present plots only for the first baseline, we note that similar trends also hold for the second one.

The Conv-TasNet-based system provides a stronger baseline as results obtained with it tend to be better. On the other hand, the mask-predictor baseline is more conventional and well-established. We note that trends we see in our results may differ for other systems that target a different goal. For reference, we provide two general system for all conditions not tailored towards a specific condition.

%%%%%%%%%%%%%%%%%%%%%%%%%
\section{Conclusions}
\label{sec:conclusions}
%%%%%%%%%%%%%%%%%%%%%%%%%
Studies in text-independent multi-channel SV in far-field conditions utilize various datasets \cite{Cai2019, Yang2019_mvdr_and_xvec, Taherian2019, Taherian2020, kataria21b_interspeech}, which makes the comparison difficult. We have reviewed recent data collecting or re-assembling activities. We pointed out reasons why they were not practical for the task of interest. Motivated by the inconsistency and lack of common data, we introduced the MultiSV corpus enabling both training and evaluation of systems dealing with noisy and reverberant speech. The training part comes with speaker labels, clean (and reverberant) speech, and noise references. Therefore, it is suitable also for (multi-channel) speech enhancement, denoising, and dereverberation. We provided four retransmitted and one simulated conditions, each having development and evaluation parts. File lists, as well as scripts for preparing MultiSV from scratch, have been made available at \url{https://github.com/Lamomal/MultiSV}.

Apart from preparing the data, we trained and evaluated two baselines. Both comprise beamforming and embedding extractors. They differ in terms of SCM estimation for beamforming: based on a mask predictor or Conv-TasNet. We made sure to provide statistically reliable baseline results for future reference and comparison.

\bibliographystyle{IEEEbib}
\bibliography{biblist}

\end{document}